\documentclass[epj,final]{svjour}
\usepackage{graphicx}
\usepackage[dvips]{color}
\usepackage{colordvi}
\DeclareMathAccent{\pol}{\mathord}{letters}{"7E}
\newcommand{\he}{$^3$\rm{He}}
\newcommand{\hepol}{$\pol{^3\rm{He}}$}
\newcommand{\heep}{$\pol{^3\rm{He}}$($\pol{e},e'p)$}
\newcommand{\heen}{$\pol{^3\rm{He}}$($\pol{e},e'n)$}
\newcommand{\hee}{$\pol{^3\rm{He}}$($\pol{e},e')$}

\newcommand{\BB}{$\pol{^3\rm{He}}$($\pol{e},e'p$)$d$}
\newcommand{\BBB}{$\pol{^3\rm{He}}$($\pol{e},e'p$)$np$}
\newcommand{\bfBB}{$\pol{^3\rm\bf{He}}$($\pol{\bf{e}}{\bf,e'p}${\bf)}${\bf d}$}
\newcommand{\bfBBB}{$\pol{^3\rm\bf{He}}$($\pol{\bf{e}}{\bf,e'p}${\bf)}${\bf np}$}
\newcommand{\Apar}{$A_{\parallel}$}
\newcommand{\Aperp}{$A_{\perp}$}
\newcommand{\gen}{$G_{\rm{en}}$}
\newcommand{\gmn}{$G_{\rm{mn}}$}
\newcommand{\emiss}{$E_{\rm m}$}

\newcommand{\et}{{\em et al.}}
\begin{document} 
\title{Measurement of the asymmetries in \bfBB\ and \bfBBB}  
\author{The A1 Collaboration \\\\
P.~Achenbach\inst{1} \and
D.~Baumann\inst{1}\and
R.~B{\"o}hm\inst{1}\and
B.~Boillat\inst{2}\and
D.~Bosnar\inst{3}\and 
C.~Carasco\inst{2}\and
M.~Ding\inst{1}\and
M.O.~Distler\inst{1}\and
J.~Friedrich\inst{1}\and
W.~Gl{\"o}ckle\inst{4}\and
J. Golak\inst{5}\and
Y.~Goussev\inst{1}$^{,}$ \inst{7}\thanks{Now at: PNPI RAS Gatchina, Leningrad district 188300, Russia}\and
P.~Grabmayr\inst{6}\and
W.~Heil\inst{7}\and
A.~H{\"u}gli\inst{2}\and
P.~Jennewein\inst{1}\and
G.~Jover Ma\~nas\inst{1}\and
J.~Jourdan\inst{2}\and
H.~Kamada\inst{8}\and
T.~Klechneva\inst{2}\and
B.~Krusche\inst{2}\and
K.W.~Krygier\inst{1}\and 
J.G.~Llongo\inst{1}\and
M.~Lloyd\inst{1}\and
M.~Makek\inst{3}\and
H.~Merkel\inst{1}\and
C.~Micheli\inst{6}\and
U.~M\"uller\inst{1}\and
A.~Nogga\inst{9}\and
R.~Neuhausen\inst{1}\and
Ch.~Normand\inst{2}\and
L.~Nungesser\inst{1}\and
A.~Ott\inst{2}$^{,}$ \inst{6}\and
E.~Otten\inst{7}\and
F.~Parpan\inst{2}\and
R.~P\'erez Benito\inst{1}\thanks{Now at: II. Phys. Institut, Justus-Liebig-Universit\"at Gie{\ss}en, Heinrich-Buff-Ring 16, 35392 Gie{\ss}en, Germany}\and
M.~Potokar\inst{10}\and
D.~Rohe\inst{2}\and
D.~Rudersdorf\inst{7}\and
J.~Schmiedeskamp\inst{7}\thanks{Now at: MPI for Polymer Research, Ackermannweg 10, 55128 Mainz, Germany}\and
M.~Seimetz\inst{1}\thanks{Now at: DAPNIA/SPhN, CEA Saclay, 
91191 Gif sur Yvette Cedex, France}\and
I.~Sick\inst{2}\and
S.~\v{S}irca\inst{10}\and
R.~Skibi\'nski\inst{5}\and
S.~Stave\inst{11}\and
G.~Testa\inst{2}\and
R.~Trojer\inst{2}\and
Th.~Walcher\inst{1}\and
M.~Weis\inst{1}\and
H.~Wita{\l}a\inst{5}\and
H.~W\"ohrle\inst{2}
}

\institute{Institut f\"ur Kernphysik, Johannes Gutenberg-Universit\"at Mainz, 
Becherweg 45, 55099 Mainz, Germany \and
Departement f\"ur Physik und Astronomie, Universit\"at Basel, Klingelbergstr.82, 4056 Basel, Switzerland \and
Department of Physics, University of Zagreb, Bijeni\v{c}ka c.~32, P.P.~331, 
10002 Zagreb, Croatia \and
Institut f\"ur Theoretische Physik II, Ruhr--Universit\"at Bochum,  Universit\"atsstr. 150, 44780 Bochum, Germany \and
M. Smoluchowski Institute of Physics, Jagiellonian University, Reymonta 4, 30059 Krak\'ow, Poland \and
Physikalisches Institut, Universit\"at T\"ubingen,  Auf der Morgenstelle 14,
72076 T\"ubingen, Germany \and
Institut f\"ur Physik, Johannes Gutenberg--Universit\"at, Staudingerweg 7, 55099 Mainz, Germany \and
Department of Physics, Faculty of Engineering, Kyushu Institute of Technology, 
1-1 Sensuicho, Kitakyushu 804-8550, Japan \and
Forschungszentrum J\"ulich, Institut f\"ur Theoretische Kernphysik, D-52425 J\"ulich, Germany \and
Institute Jo\v{ze}f Stefan, University of Ljubljana, Jamova 39, 1000 Ljubljana, 
Slovenia \and
Laboratory for Nuclear Science, Massachusetts Institute of Technology, Cambridge, MA, USA}

\mail{D. Rohe, e-mail: Daniela.Rohe@unibas.ch}

\authorrunning{P. Achenbach \emph{et al.}}
\titlerunning{Measurement of the asymmetries in \bfBB\ and \bfBBB} 
\date{\today}

\abstract{The electron-target-asymmetries \Apar\ and \Aperp\ with target spin parallel and perpendicular to the momentum transfer ${\bf q}$ were measured for both  the two-- and three-body breakup of \he\ in the \heep-reaction. Polarized electrons were scattered off polarized \he\ in the quasielastic regime in parallel kinematics with the scattered electron and the knocked-out proton detected using the Three-Spectrometer-Facility at MAMI. The results are compared to Faddeev calculations which take into account Final State Interactions as well as Meson Exchange Currents. The experiment confirms the prediction of a large effect of Final State Interactions in the asymmetry of the three-body breakup and of an almost negligible one for the two-body breakup.
}

\maketitle 

\section{Introduction}\label{intro}
Three-nucleon systems are good testing cases for our understanding of the nuclear ground state and reaction mechanisms. The number of nucleons and their density is large enough to exhibit all important features of an $(e,e'p)$-reaction  on complex nuclei. On the other hand it is small enough to allow exact calculations. One of the first works investigating the influence of Final State Interactions (FSI) and Meson Exchange Currents (MEC) on the spin observables in 3-body systems uses a diagrammatic approach \cite{Laget91,Laget92} which already predicts the basic features of the reaction mechanism.  Nowadays calculations based on the non-relativistic Faddeev equation use modern nu\-cleon-nucleon potentials and have reached a high degree of sophistication. They are able to treat FSI as well as MEC \cite{Golak02,Schulze93}. Recently, calculations  using relativistic kinematics and a relativistic current operator, but treating  the FSI to first order only, became available \cite{Carasco03}. Three-nucleon systems are also the natural place to study three-body forces which have attracted considerable attention \cite{bound_state,elasticNd,3Nbreakup,ChPT}. While the need for a three-body force in electron-induced break-up observables is not yet clear, it requires a careful study, since its effect might be obscured {\it e.g.} by  an incomplete treatment of the nuclear current operator and/or relativistic effects. The analysis of the photodisintegration reactions \cite{Skibinski1,Skibinski2} suggests that in general the three-body force plays a significant role. 

With the availability of highly polarized \he\ of sufficient density and the delivery of polarized continuous electron beams of high intensity, spin-dependent quantities can be studied, which show a large sensitivity to the underlying nuclear structure and reaction mechanism. Since in \he\ the protons reside with high probability in the $S$-state, the spin of \hepol\ is essentially carried by the neutron \cite{Blank84}. This characteristic of the \he-spin structure can be best exploited in the quasielastic reaction \heen\ with restriction to small missing momenta as well as in inclusive \hee\ near the top of the quasielastic peak. In such kinematics the \hepol-target has been used extensively as polarized neutron target  to measure the magnetic \cite{Gao94,Xu00,Xu03} and electric \cite{Meyerhoff94,Becker99,Golak01,Rohe99,Bermuth03} form factors of the neutron, \gmn\ and \gen.  The interference term of the electric and magnetic scattering amplitudes leads to an asymmetry from which \gen\ is obtained with good precision, but it has to be corrected for  FSI and MEC contributions. In particular at small momentum transfer $Q^2$ the effect of FSI is significant. As we shall see in this paper different aspects of the underlying spin structure of \hepol\ are revealed in the reaction channels \BB\ and \BBB. In particular when considering the effect of FSI the simple picture can change dramatically.

A detailed knowledge of the ground state wave function and a precise treatment of the reaction process is furthermore of importance for experiments which aim to extract the neutron spin structure functions from \hepol\ ({\it e.g.} Gerasimov-Drell-Hearn sum rule). The present study provides an important test of the reliability of the necessary theoretical description of the three-body system.

We have measured the electron-target-asymmetries \Apar\ and \Aperp\ with target spin parallel and perpendicular to the momentum transfer ${\bf q}$ on top of the quasielastic peak in the reactions \BB\ and \BBB. The knocked-out proton was detected in the direction of ${\bf q}$ (parallel kinematics). Here the theory predicts a large influence of FSI on the three-body breakup (3BB) but only a small effect for the two-body breakup (2BB). Contrary to earlier inclusive measurements \cite{Xiong01} our experiment allows a clear separation of the 2BB and 3BB. The data were taken  together with a study \cite{Proposal00} of the $D$-state admixture in the ground state of \he, which contributes to the high-momentum-components of the wave function \cite{Schulze93,Ericson87,Blank84}.

In section 2 we explain the experimental setup. Section 3 is devoted to the
analysis of the data. While the peak of the 2BB can be separated from the
3BB by a cut in the missing energy, the 3BB  is affected by the radiation
tail of the 2BB; the separation has been performed by a Monte Carlo
simulation. In section 4 we compare the measured data to our theoretical
description.

\section{Experimental setup} \label{setup}
The experiment was carried out at the Three-Spectrometer-Facility of the A1 collaboration at MAMI \cite{Blo98}. The polarization of the incoming electron beam, $P_{\rm e}$, produced by photoelectron emission from a stressed-layer GaAsP crystal \cite{Aule97}, was measured twice a day with a M{\o}ller polarimeter located a few meters upstream of the target. $P_{\rm e}$ was constant within the statistical error bars of the individual measurements in each of the two separate run times. In the second beam time period an increased $P_{\rm e}$ of  81~\% was achieved compared to 73~\% in the first part; the luminosity-weighted average over the two run times of three weeks in total amounted to (76.21 $\pm$ 0.16 (stat.) $\pm$ 1.3 (syst.)) \%. The current was held constant at 10~$\mu$A, the maximum tolerated by the polarized target system. 

Details of the setup for the polarized target were given earlier \cite{Rohe99,Bermuth03}. \he\ is polarized by the technique of meta\-stable optical pumping at $\approx$~1~mbar and subsequently compressed to $\approx$ 5 bar by a titanium piston compressor \cite{Otten04}. Polarizations of up to 70~\% were reached. The gas was filled into a glass container consisting of a spherical part (diameter 80~mm) with cylindrical extensions on both sides. These are closed by end caps of 25 $\mu$m thick Cu-foils which serve as entry and exit windows for the electron beam. The windows are outside of the acceptance of the spectrometers to reduce background. In addition, the target area is shielded with lead bricks. The polarization of \he\ is maintained by a homogeneous magnetic field of 4~G. It is provided by three coils wrapped around a box of $\mu$-metal and iron plates which serve as shield against the stray field of the magnetic spectrometers. With a careful preparation of the glass container, covered by a layer of cesium, relaxation times of 70 to 80~h were achieved at a pressure of 2~bar. Due to additional relaxation mechanisms like collisions of the \he\ atoms among each other as well as the ionization of \he\ by the electron beam, the lifetime of the polarization is effectively reduced. With the known values for the influence of the electron beam and the pressure on the relaxation time \cite{Chupp92,Newbury93} one computes for our conditions an in-beam relaxation time of $\approx$~45~h. The in-beam relaxation times measured in nine different target cells varied between 20 and 40~h. They were changed twice a day. Surprisingly it was observed that the relaxation time increased with repeated use in the electron beam. This effect is not entirely understood yet. The reason  might be outgassing of the glue, used for the copper foil attachment. The outgassing could be initiated by the  electron beam which would lead to a contamination of the \he\ gas and to a reduction of the relaxation time due to paramagnetic centers. After several refills outgassing is reduced.

The polarization of \he\ was monitored with Adiabatic Fast Passage (AFP) using the technique described in ref. \cite{Wilms97} which measures the magnetic field of the oriented spins. The main systematic error of AFP is due to the distance measurement between the magnetometer and the center of the spherical target. The method was compared to and found to be in agreement with an absolute polarization measurement performed at the TRIGA reactor of the Mainz University \cite{Baess}. This method exploits the polarization-dependence of the neutron-flux through the \he-container. Since the AFP-technique destroys part of the polarization ($\approx$ 0.1 -- 0.2~\%) and since it cannot be used during data-taking due to spin-flipping, it is used only about once in 4 h. Therefore Nuclear Magnetic Resonance (NMR) monitored continuously ($\approx$ every 10 min) the relative polarization and served mainly as online control of the polarization. The systematic error of the absolute polarization is estimated to be 4~\% and the uncertainty in the relaxation time is 2~h. Averaged over the beam time the target polarization, $P_{\rm T}$, was (49.8  $\pm$ 0.3 (stat.) $\pm$ 2 (syst.)) \%.

The scattered electron and the knocked-out proton from the reaction \heep\ were detected in the high-resolution magnetic spectrometers A and B, respectively. The kinematic setting is given in table \ref{kin}. The $Q^2$ varied over the detector acceptance between 0.25 and 0.4 (GeV/c)$^2$ and p'$_{\rm e}$ between 540 and 575~MeV/c. The missing momentum $|\vec{p_m}|$ = $| \vec{q} - \vec{p_p}|$ is less than 120~MeV/c and on average 40~MeV/c. In the same kinematics and setup data with H$_2$ were taken for checks (see below). 

The electron helicity was randomly flipped every second. The target spin direction was changed after every hour of data taking, cycling through the spin orientations parallel, perpendicular, antiparallel and antiperpendicular relative to the momentum transfer. This reduces systematic errors considerably.

\begin{table}[t]
\caption{\label{kin} The kinematical setting. E: beam energy, p'$_{\rm e}$ (p$_{\rm p}$): central momentum of spectrometer A (B), $\theta_{\rm e}$ ($\theta_{\rm p}$): central angle of spectrometer A (B). The given four-momentum transfer $Q^2$ is averaged over the detector acceptance.}
\begin{tabular}{llllll}
\hline\noalign{\smallskip}
Q$^2$ & E & p'$_{\rm e}$ & $\theta_{\rm e}$ & p$_{\rm p}$ & $\theta_{\rm p}$ \\
(GeV/c)$^2$ & MeV & MeV/c & deg. & MeV/c & deg. \\
\noalign{\smallskip}\hline\noalign{\smallskip}
0.31  & 735 & 600 & 53 & 600 & 48.5 \\
\noalign{\smallskip}\hline
\end{tabular}
\end{table}

\begin{figure}[t]
\begin{center}
\includegraphics[width=6cm,clip]{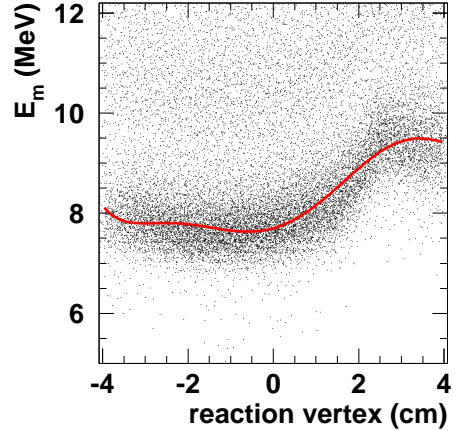}
\end{center}
\caption{\label{Emtarz}Missing energy, as calculated by Eq. \ref{asymexp}, against the reconstructed reaction vertex position along the beam axis. Positive values are downstream. The thick line is a fit to the \emiss\ distribution belonging to the 2BB.}
\end{figure}
\section{Data analysis} \label{ana}
\begin{figure}[t]
\begin{center}
\includegraphics[width=8cm,clip]{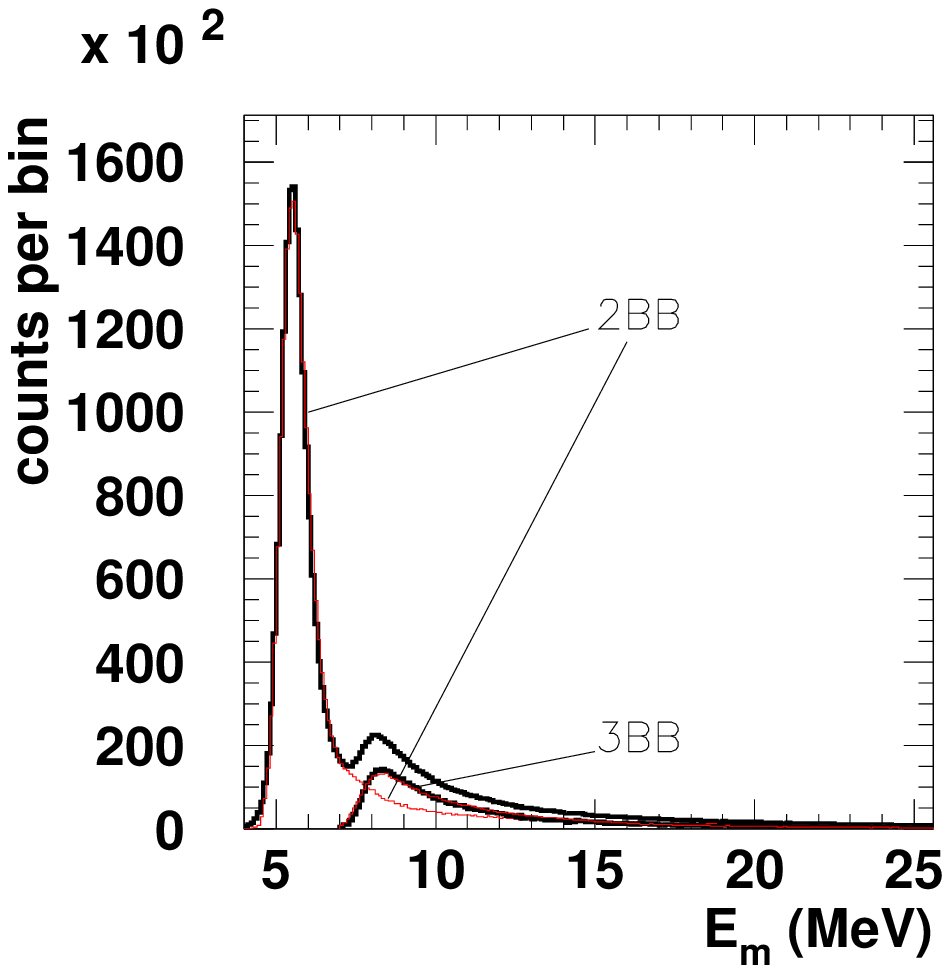}
\end{center}
\caption{\label{Emsimulin}Experimental \emiss\ distribution (thick line 2BB) and the
simulation of the 2BB (thin red line). The difference is shown as thick black line 3BB.}
\begin{center}
\includegraphics[width=8cm,clip]{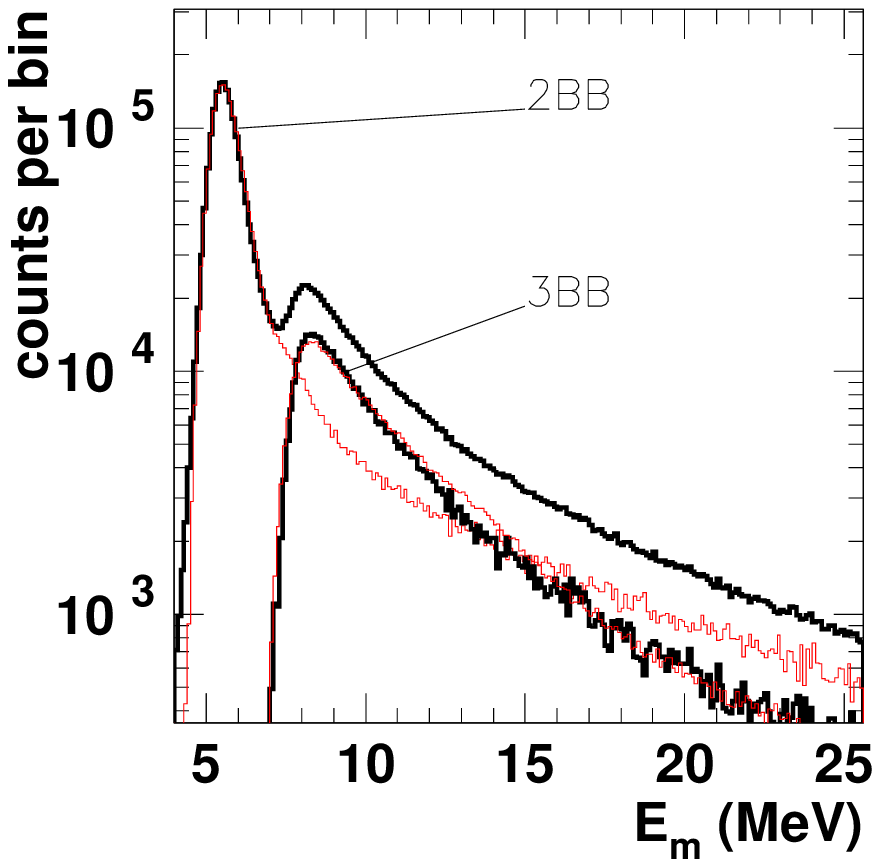}
\end{center}
\caption{\label{Emsimulog} Same as fig. \ref{Emsimulin} on a logarithmic scale. In addition the simulation of the 3BB is shown in red.}
\end{figure}

From the measured kinematic variables, the missing energy is reconstructed
according to
\begin{equation} \label{Em}
E_m = E - E_e - T_p - T_R .
\end{equation}
Here, $E$ ($E_e$) is the initial (final) electron energy and $T_p$ is the kinetic energy of the outgoing proton.  $T_R$ is the kinetic energy of the (undetected) recoiling (A-1)-system, which is reconstructed from the missing momentum under the assumption of 2BB; the error made by also using this for the 3BB is of minor importance due to the smallness of the missing momentum in the chosen kinematics. Due to energy losses of the outgoing electron and proton, especially in the 2~mm thick glass wall of the target cell, \emiss\ is shifted to larger values. In addition there exists a dependence of \emiss\ on the reconstructed reaction vertex position along the beam axis (see fig. \ref{Emtarz}). It reflects the fact that the thickness of the target wall is not constant. This pattern differs from cell to cell. A maximum of the energy loss between 2 and 3 cm downstream from the center of the target cell is common to most of the target containers. It is due to an increase of the glass thickness at the position where the cylindrical extension is attached to the spherical part. The spectrum shown in fig. \ref{Emtarz} was fitted for each target cell individually to correct \emiss\ for this dependence. The corrected \emiss\ distribution is shown as thick black line in fig. \ref{Emsimulin}. No indication of background is found in the \emiss\ spectrum. The resolution is limited mainly by the properties of the target cell and not by the resolution of the spectrometers.

The FWHM of 1~MeV allows a clear separation of the the \emiss-regions
where only 2BB or 2BB and 3BB contribute. The \emiss-region from 4.0 to 6.5 MeV is interpreted as pure 2BB. This cut was chosen to avoid any contribution from the 3BB-channel (starting at 7.7 MeV) considering the experimental \emiss\ resolution. In agreement with ref. \cite{Flori99}, the yield of the 3BB is negligible beyond 25 MeV. Therefore the cut for the 3BB-channel was made from 7.5 to 25.5 MeV in the \emiss\ spectrum. Because the 3BB resides on the radiation tail of the 2BB, the latter has to be accounted for in the analysis of the 3BB-region of the measured spectrum. To this end, the tail was calculated in a Monte Carlo simulation which accounts for internal and external bremsstrahlung, ionization loss and experimental energy resolution adjusted to the experimental distribution. In the simulation the detector acceptances were taken into account. The quality of the calculation was checked on H$_2$-data where the tail region is not obscured by 3BB. Good agreement between calculation and measured data was found. In the simulation, the inhomogeneous thickness of the glass container was taken into account, but a small dependence of \emiss\ on the reaction vertex remained. The resulting uncertainty was accounted
for in the systematic error of the 3BB data.

The simulation for \he\ starts from the initial momentum distribution of the bound proton and subsequently uses the kinematical relations valid in PWIA. 
Two momentum distributions were used in the Monte Carlo simulation. One stems from a fit to the experimental results of ref. \cite{Flori99}. The other results from a theoretical calculation using the Reid soft-core interaction \cite{Dieperink76}. As expected the dependence of the simulated \emiss\ spectrum on the momentum distribution in \he\ is small. In order to gain confidence in the subtraction of the 2BB and thus in the determination of the 3BB-contribution, we also performed a Monte
Carlo calculation of the 3BB-channel. For this simulation we took
directly the theoretical \emiss\ distribution from ref. \cite{Dieperink76}.

Fig. \ref{Emsimulin} shows the measured \emiss\ spectrum together with the calculated
2BB-channel. Also shown is the result after subtracting the simulated
2BB contribution from the experimental data, which we interpret as
3BB-channel. For a better display of the radiation tail and of the
3BB, the same plot is shown with logarithmic scale in fig. \ref{Emsimulog}. Here
we also show the simulation of the 3BB, which, apart from a small excess
around \emiss\ = 14 MeV, agrees very well with the 3BB data. The small
deviation can be traced back to a little bump in the theoretical
spectral function of ref. \cite{Dieperink76}; it should be of no importance for this analysis.

The ratio of the Monte Carlo simulation of the 2BB to the experimental data in the region of the 3BB is denoted by $a_{\rm 23}$. For the region 7.5 $<$ \emiss\ $<$ 25.5 MeV it amounts to  $a_{\rm 23}$ = 0.434 $\pm$ 0.002 (stat.) $\pm$ 0.015 (sys.). The systematic error was estimated from simulations using different ingredients (in particular different inhomogeneous glass thicknesses of the container) and from the sensitivity to cuts like {\it e.g.} the reconstructed reaction vertex. 

For the four target spin angles the experimental asymmetry is obtained via
\begin{equation} \label{asymexp}
A_{exp} = \frac{N^+/L^+ - N^-/L^-}{N^+/L^+ + N^-/L^-},
\end{equation}
 where $L^+$ ($L^-$) are the integrated charge and  $N^+$ ($N^-$) the number of events for positive (negative) electron helicity within the above limits for the 2BB- and 3BB-channels. No background needs to be subtracted: It is less than 0.3~\% in the coincidence time spectrum and vanishes entirely in the \emiss\ spectrum within the used limits. The charge asymmetry $(L^+ - L^-)/(L^+ + L^-)$ was constant during
the measurement and amounted to 0.2~\%; it has no influence on the accuracy of the extracted asymmetries. 

The average direction of the momentum transfer does not exactly match the reference direction of the target spin, i.e. the one for the parallel asymmetry. Therefore,  $A_{exp}$ contains contributions from the perpendicular and parallel  asymmetry, respectively: 
\begin{equation}\label{asymthq}
A_{exp} = A_{\parallel}  \cos\Delta\theta + A_{\perp}  \sin\Delta\theta .
\end{equation}
Here $\Delta\theta$ is the difference between the target spin angle, $\theta_{\rm T}$, and the angle of the momentum transfer, $\theta_{\rm q}$. The mean angle $\theta_{\rm q}$ was constructed from measured quantities for the events in the 2BB-peak of the \emiss\ distribution. The target spin angle $\theta_{\rm T}$ was measured with a magnetometer, located 6~cm below the center of the target cell, with a systematic uncertainty of 0.2${^\circ}$. Due to gradients in the magnetic field of the target box small deviations from the actual value at the position of the target cell can occur. This has been corrected for by using calibration data taken during the setup of the experiment. After each change of the magnetic field $\theta_T$ is remeasured. The luminosity-weighted averages of the target spin angles amounted to 48.6$^{\circ}$, 138.5$^{\circ}$, 228.6$^{\circ}$ and 318.5$^{\circ}$ relative to the beam direction. Relative to the central photon direction as given in Table \ref{kin} for parallel kinematics these angles correspond to parallel, perpendicular, antiparallel and antiperpendicular spin directions, respectively. The maximum deviation from the parallel and antiparallel directions, respectively, was 0.4$^\circ$. This leads only to a small correction. 

In order to compare the measured with the theoretical asymmetry, $A_{\rm exp}$ has to be normalized to the electron and target polarizations:
\begin{equation}\label{asym_corr}
A = \frac{1}{P_e} \frac{1}{P_T} A_{exp} .
\end{equation}
While the data with 4.0~MeV $<$ \emiss\ $<$ 6.5~MeV uniquely define the 2BB-events
and thus determine the asymmetry in this channel, $A_{\rm 2BB}$, the region 7.5~MeV $<$ \emiss\ $<$ 25.5~MeV is fed by the 3BB and by the radiation tail of the 2BB. The asymmetry in this \emiss-region is named $A_{\rm 2+3BB}$. In order to extract the asymmetry $A_{\rm 3BB}$ for the 3BB-channel, $A_{\rm 2+3BB}$ has to be corrected for the 2BB-contribution.  This is achieved via
\begin{equation} \label{corr_2BB}
A_{3BB} = \frac{A_{2+3BB} - A_{2BB} \, a_{23}}{1 - a_{23}} .
\end{equation}

\section{Results and discussion} \label{res}
The measured asymmetries are summarized in table \ref{asyms} together with
the statistical and systematic (first and second value in parentheses) errors. They are given for the four spin angles separately and also as mean over the (anti)parallel ($\parallel$) and (anti)perpendicular ($\perp$) directions, respectively, with appropriately accounting for changes in the sign. The systematic errors from the electron and target polarization as well as from the measurement of the target spin direction were taken into account. Because the values of the asymmetry for parallel and perpendicular target spin are quite similar in size for the 2BB and 3BB, respectively, the sensitivity to the target spin direction is equally small for both asymmetries. The main uncertainty in the extraction of the 3BB asymmetry  comes from $a_{\rm 23}$.  

\begin{table}[t]
\caption{\label{asyms} Asymmetries for the four target spin angles $\theta_{\rm T}$ in the 2BB- (\emiss : 4.0--6.5~MeV) and 3BB-region (\emiss : 7.5--25.5~MeV). A$_{\rm 3BB}$ is the asymmetry after correcting for the 2BB-contribution in the 3BB-region. In parentheses the statistical and systematic errors are given. In addition the averages for parallel ($\overline{\parallel}$) and perpendicular ($\overline{\perp}$) target spin orientations are presented.}
\begin{tabular}{rrrr}
\hline\noalign{\smallskip}
$\theta_{\rm T}$ & $A_{\rm 2BB}$& $A_{\rm 2+3BB}$& $A_{\rm 3BB}$\\
\noalign{\smallskip}\hline\noalign{\smallskip}
48.5   &   0.1141(41,48)  & $-$0.125(5)& $-$0.3089(98,151)  \\
138.5  &$-$0.1424(44,62)  &    0.116(6)&    0.3142(104,158)  \\
228.5  &$-$0.1345(46,59)  &    0.120(6)&    0.3153(110,158)  \\
318.5  &   0.1284(48,58)  & $-$0.137(6)& $-$0.3406(116,172)  \\
$\overline{\parallel}$&   0.1231(30,53)  & $-$0.123(4)& $-$0.3117(74,154)  \\
$\overline{\perp}$ &$-$0.1360(32,60)  &    0.125(4)&    0.3259(78,164)  \\
\noalign{\smallskip}\hline
\end{tabular}
\end{table}

\begin{figure}[t]
\begin{center}
\includegraphics[width=8cm,clip]{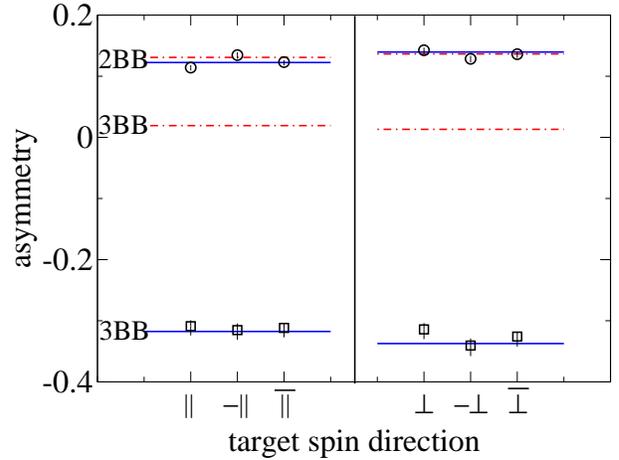}
\end{center}
\caption{\label{asym_all}Comparison of the data shown in table \ref{asyms} to the theoretical calculation for the 2BB and 3BB for the four target spin directions (anti)parallel ($\parallel$, --$\parallel$; left panel) and (anti)perpendicular ($\perp$, --$\perp$; right panel). In addition the combined sum for the parallel and perpendicular position is shown ($\overline{\parallel}$ and $\overline{\perp}$, respectively). To facilitate the comparison, all 2BB (3BB)data are shown with positive (negative) sign. PWIA: dot-dashed lines. Full calculation including FSI and MEC: solid lines. Statistical errors point up, systematic uncertainties point down. For the 2BB the size of the error bars is smaller than the symbols.}
\end{figure}

In fig. \ref{asym_all}, the measured asymmetries are compared to our corresponding calculations using the AV18 potential. The variation of the calculated asymmetry with the proton scattering angle is negligible over the range covered by spectrometer B (horizontal: $\pm$~20 mrad, vertical: $\pm$ 70~mrad). However, the asymmetry shows large variations with the electron angle (horizontal: $\pm$ 100~mrad, vertical: $\pm$ 70~mrad) and the final electron momentum in both channels ($\Delta p$ = 40~MeV/c). Therefore asymmetries for ten different kinematics were calculated and averaged over the detector acceptance. The result using PWIA only is shown by the dashed-dotted line. The calculation including FSI accounts for the interaction between the two spectator nucleons (rescattering term) as well as between the hit nucleon and the spectator(s) (direct FSI). It is shown by the solid line. The effect of MEC is negligible in this kinematics. The data integrated over the total detector acceptance are in good agreement with the calculation including FSI. 

The calculation shows that the FSI contribution is small in the 2BB
while it is large in 3BB. This suggests that the main contribution of FSI results from the rescattering term which does not exist in the 2BB, and not from direct FSI. The same conclusion was already drawn in another context in refs. \cite{Carasco03} and \cite{Bermuth03}. 

Since the 2BB is quantitatively described by the PWIA calculation, it can be interpreted as follows: The polarized \hepol\ can
be described by the disintegration into a deuteron and a proton, the
spins of which couple to the \he-spin according to
\begin{eqnarray} \label{coup}
 |(1,1/2)\, 1/2,+1/2> &=& \sqrt{\frac{2}{3}} |1,1> |1/2,-1/2> \nonumber\\
            &  & - \sqrt{\frac{1}{3}} |1,0> |1/2,+1/2> .
\end{eqnarray}
The first value in the ket-symbols on the right hand side of eq. \ref{coup} is the spin $S$, the second one the projection on the quantization axis $M_S$. On the left hand side the subsystem in \he\ consisting of two nucleons couples first to spin 1 and then with the remaining nucleon of spin 1/2 to the \he-spin $S = 1/2, M_S = +1/2$.  From eq. \ref{coup} one derives that for the 2BB-channel a 100\%-polarized \hepol\ target constitutes a proton target which is 33\% polarized in the direction opposite to the \he-target. This value is also confirmed by the spin asymmetry in the momentum distribution at small nucleon momenta obtained by a Faddeev calculation shown in Fig. 2 of ref. \cite{Milner96}. For the given kinematics, the asymmetry on a polarized proton is 39.2~\% for the parallel and --41.4~\% for the perpendicular spin direction. Combining this with the degree of proton polarization gives our result for the 2BB of polarized \hepol, namely 13.1~\% and --13.8~\%, respectively. Therefore, in the 2BB-channel \hepol\ can be regarded as a proton target with a polarization reduced by a factor 1/3 with respect to that of \hepol.

In PWIA the asymmetry is almost zero for the 3BB which reflects the fact that the two protons, which are dominantly in the $S$-state and thus have opposite spin orientation, now contribute equally to the knock-out reaction. The inclusion of FSI, however, leads to an asymmetry, which is larger and opposite in sign compared to the 2BB. This effect was already predicted in ref. \cite{Nag98}, where it was explained by the large difference between the singlet and the triplet pn-interaction at low energies of the spectators. One could naively expect that the np t-matrix is dominated by the singlet interaction
and thus the spins of the neutron and one of the protons are opposite to each other. This would lead to the effect that the knocked-out proton carries up to 100 \% the \hepol-spin. However this is not the case, since also the triplet np t-matrix contributes. Therefore the response functions depend on different combinations of spin orientations of the np spectator pair and the knocked out proton. As a consequence, the asymmetries carry information on the rescattering mechanism and the \he\ state, which does not drop out. This does not allow to consider this reaction channel as a scattering process on a free polarized proton as it was argued in ref. \cite{Nag98}. In this reference the authors conclude that the proton-polarization in the 2BB- and 3BB-channels should be equal in size with only opposite sign. However, the experimental and theoretical results presented in this paper do not support this statement. The same conclusion can also be drawn from the coupling scheme described in eq. \ref{coup}.

Furthermore, the 3BB-region was split into five equal bins from 7.5 to 25.5 MeV. Effects from radiative processes were unfolded in two steps. First the asymmetry in the 3BB-channel was obtained by accounting for the radiation tail from the 2BB-channel using Eq. \ref{corr_2BB} but applying it for each bin separately. Caused by the different shapes of the radiation tail of the 2BB and the distribution of the 3BB the factor $a_{\rm 23}$ is different for each bin. Second,  due to radiative processes within the 3BB-channel events are shifted from one bin to subsequent bins at higher \emiss. This leads  to a redistribution of the number of events and subsequently to a mixing of the asymmetries between different \emiss\ bins. To account for this effect the theoretical asymmetry as a function of \emiss\ was fed into the Monte Carlo simulation and the change of the asymmetry in each bin without and with radiative processes was recorded. This factor was used to correct the experimental asymmetries in each of the five bins.

\begin{figure}[t]
\begin{center}
\includegraphics[width=8cm,clip]{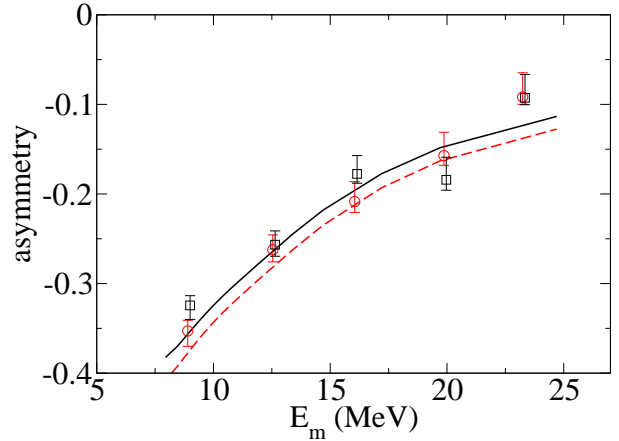}
\end{center}
\caption{\label{Embins}Parallel and perpendicular asymmetry as a function of \emiss\ in the 3BB-channel. \Apar: data (squares), theory (solid line). \Aperp: data (circles), theory (dashed line). Statistical errors point up, systematic uncertainties point down. The squares are (artificially) shifted in \emiss\ by +0.1 MeV to avoid overlapping data points.}
\end{figure}

The result for the parallel (squares) and perpendicular (circles) asymmetry is compared to theory in fig. \ref{Embins}. Within the error bars the experimental and theoretical results are in good agreement. Also here MEC do not play a role. It was tested that a calculation using the CDBonn potential instead of the AV18 did not lead to significantly different results. In addition, the Urbana IX three-nucleon force was included for the first time in our calculation for the \BB\ and \BBB\ processes. The calculation performed for the central kinematics of this experiment shows a difference of less than $-$0.01 in the asymmetry at large \emiss\ compared to the result without three-nucleon force. 

\section{Summary} \label{conc}
Using two high resolution spectrometers in coincidence it was possible to separate the regions of the 2BB- and 3BB-channel in the reaction \heep\ on top of the quasielastic peak in parallel kinematics. Mainly caused by the wall of the target container the FWHM of the \emiss\ distribution in the 2BB-channel is broadened to 1~MeV. The radiative tail of the 2BB contributes more than 40\% in the \emiss-region of the 3BB. A Monte Carlo simulation of the 2BB-channel was used to extract the asymmetry for the 3BB. Due to the large opposite asymmetries for the 2BB and 3BB some sensitivity of the 3BB asymmetry to the input parameters of the simulation is present. It contributes the largest part to the systematic error of the extracted 3BB asymmetry. As a check of the analysis the \emiss\ spectrum of the 3BB was also simulated and found in good agreement with the measurement except for a small bump around 14~MeV. 

The measured parallel and perpendicular asymmetries of the 2BB and 3BB are in excellent agreement with our Faddeev calculations. This holds also for the more detailed comparison in the 3BB-channel in five \emiss-bins. This confirms that our theory is a reliable tool for the description of reactions using \hepol.

The calculation shows that MEC-contributions are negligible in both channels and the influence of the three-nucleon force is small. In our kinematics, the asymmetry in the 2BB-channel is governed by the impulse-approxima\-tion, i.e. FSI-contributions are negligible. In this channel \hepol\ can be regarded as a proton target with a polarization being given by the (Clebsch-Gordon-)weights of the coupling of the proton and the deuteron spins to the \he-spin; this results in a 30\%-polarization of the proton opposite to that of the \hepol. In contrast, the asymmetry in the 3BB-channel is large and dominated by both, the singlet and the triplet interaction between the spectators (rescattering term). This leads to an asymmetry which cannot be considered as resulting from scattering on a polarized proton. There is still dependence on the \hepol-spin structure and the rescattering mechanism. The underlying reaction and coupling mechanisms leading to the asymmetries for the \BB\ and \BBB\ reaction channels will be further theoretically investigated in a forthcoming paper.

\section*{Acknowledgments}
We like to thank the MAMI accelerator group and the MAMI polarized beam group for the excellent beam quality combined with a continuous high polarization. This work was supported by the Sonderforschungsbe\-reich 443 of the Deutsche Forschungsgemeinschaft, the Schweizerische Nationalfonds (SNF), the DFG Graduiertenkolleg Basel-T\"ubingen GRK 683 as well as by the  Polish Committee for Scientific Research under Grant no. 2P03B00825 and by the NATO grant no. PST.CLG.978943. The numerical calculations have been performed on the Cray SV1 of the NIC in J\"ulich, Germany.

\end{document}